\documentstyle[prl,aps,amsfonts,amssymb,twocolumn,epsfig]{revtex}

\preprint{
\hfill  TH00.2
} 

\title{
Intermediate spin, Schr\"odinger cat states and nano-magnets.
} 

\author{S. E. Barnes }

\address{ Department of Physics, University of Miami, Coral Gables, 
Florida 33124 }

\begin{document}
\draft

\twocolumn[
\hsize\textwidth\columnwidth\hsize\csname @twocolumnfalse\endcsname

\date{\today} \maketitle
\begin{abstract}
Quantum tunneling of nano-magnets finds a natural description in terms 
of intermediate spin.  Periodic magnetic effects correspond to a change of 
flux by the flux quantum $\Phi_{0}$. Schr\"odinger 
cat states with different superpositions  of the applied magnetic 
field occur. The molecular magnet Fe$_{8}$ is discussed and oscillations are 
predicted for Mn$_{12}$.
\end{abstract}

\pacs{75.45+j,75.50Tt,75.60Ej}

\vskip1.0pc]

Recent experiments by Wernsdoffer and Sessoli\cite{ws} on the 
molecular magnet known as Fe$_{8}$ show fasinating behaviour as a 
function of an applied field.  This is an orthrombic easy axis 
material with one perpedicular axis harder than the other.  As a 
function of either an easy {\it or\/} hard axis field the tunnel 
splitting manifests self-similar behavior at periodic values of these 
fields.  As a function of the third component the splitting increases 
essentially exponentially.  While these results can be well reproduced 
by the exact diaginalisation of a suitable large spin model, as yet 
there is not an analytic theory which permits a clear understanding of 
the observed phenomina.  An adequate such theory should permit the 
prediction of new related effects for materials, e.g., of different 
symmetry.

\begin{figure}[b!]
\centerline{\epsfig{file=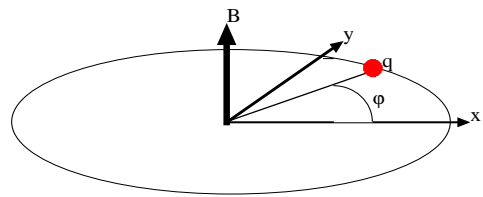,height=1.in,width=2.5in} } 
\vskip 10pt
\caption[toto]{A particle of charge $q$ is confined to a unit circle. 
A tunneling problem might be defined by adding 
a potential $V$ such that $V(\phi) = V(\phi + \pi)$. If this 
potential has minima at $\phi = 0$ and $\pi$ and the barrier between is 
higher than the zero point energy then there will be a ground doublet 
with a splitting $\Delta$. A perpendicular field exerts no force on 
the particle but does change the slitting as explained in the text.
}
\label{interspin}
\end{figure}

Invariably, in the tunnelling regime, the least action paths which 
make any significant contribution to the tunneling amplitude can be 
mapped to a circle.  The problem is then equivalent to the motion of a 
charged particle in a magetic field, Fig.~(\ref{interspin}), i.e., to 
the problem of {\it intermediate spin}.  In this Letter will be 
described an analytic theory based upon an extension of the 
intermediate spin approach developed earlier by the 
author\cite{barnes}.  In terms of this the behavior of Fe$_{8}$ is 
easily understood and, e.g., a new periodic effect can be predicted 
for the large spin model often used to describe the much 
studied\cite{thomas} Mn$_{12}$.
 
The relevant spin model for Fe$_{8}$ can written as:
\begin{eqnarray}
&{\cal H} = - (D-E)& {S_{x}}^{2} + 2 E {S_{z}}^{2} \nonumber \\
&&- E S^{2}- h S_{z} - h_{\ell} S_{x}-h_{t}S_{y},
\label{un}
\end{eqnarray}
where the positive anisotropy constants have $E<D$\cite{seb}.  The 
magnetic field $h$, parallel to the hardest axis, is equivalent to 
that in Fig.~(\ref{interspin}) and most physical quantities, including 
the tunnel splitting, must be periodic in the flux $\Phi$ produced by 
this field, this with a period $\Phi_{0}$, the flux quantum.  If the 
physical spin value $S$ is whole integer, a flux of 
$(n+{1\over2})\Phi_{0}$ corresponds to a statistical parameter 
$\alpha_{s} = 1$, modulo unity, and transmutes the model to 
half-integer spin while intermediate values $n\Phi_{0}$ reflect a 
$\alpha_{s} = 0$, modulo unity, and are whole-integer points.  
Kramer's theorem implies a zero tunnel splitting for half-integer 
points.  These facts are reflected in the quasi-periodic tunnel 
splitting $\Delta = \Delta_{0}\cos 2\pi \Phi/\Phi_{0}$ with 
$\Phi/\Phi_{0}=h/4 \hbar \sqrt{2E(D+E)}$ evident in 
experiment\cite{ws}.

There are some interesting surprises, e.g., in order to describe 
tunneling in Fe$_{8}$ it is necessary to construct Schr\"odinger cat 
states which involve different values of the applied magnetic field.  
Even more bizarre is that the other transverse field $h_{t}$ is 
equivalent to an {\it imaginary\/} flux $\Phi$ leading to a $cosh$ 
dependence of the splitting.  In general for an arbitrarly directed 
transverse field the effective flux is complex!

There are classical ground states which with $h=h_{t}=0$ lie along the 
positive and negative $x$-axes.  Tunneling occurs between these ground 
states.  Adding a the longitudinal field $h_{\ell}$,  e.g., the ground 
state for one direction can be brought to the same energy as the first 
excited state in the other direction. In this case one wave function has 
even parity and the other odd and this adds a phase shift of 
$\pi$ to the wave function. This also converts whole into 
half-integer spin, again an effect seen in experiment\cite{ws}.

These ideas permit the prediction of new effects even for the 
extensively studied molecular magnet Mn$_{12}$.  This is a tetragonal 
easy y-axis magnet but with equivalent $x$ and $y$ hard directions in 
the transverse plane.  There are {\it four\/} equivalent least action 
tunnel paths in the absence  of a field.  When a modest field is 
applied along any of the four hard directions there remain only two 
equivalent least action paths which are mapped to the intermediate 
spin circle.  Actually this problem can be solved even for\break
\begin{figure}[t!]
\centerline{\epsfig{file=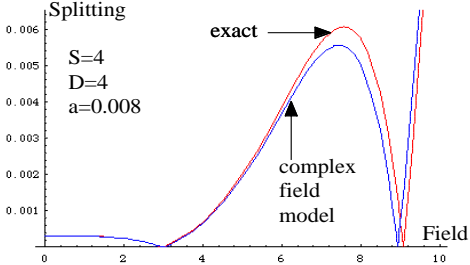,height=1.4in,width=2.5in} } 
\caption[toto]{The analytic result is $\Delta_{0}|\cosh(\pi h 
/p)\cos(\pi h /p)|$ where $ p \approx 2 S^{1/2} (12 a D^{3})^{1/4}$.  This 
corresponds to a complex magnetic field in which the real and 
imaginary parts are equal.  This exact result is indicated.  The 
difference becomes smaller with increasing $S$.}
\label{Mn}
\end{figure}
\noindent small fields by mapping the four paths to {\it two\/} 
circles and then adding the wavefunctions in much the same way as in 
the elementary theory of a two slit experiment.  These circles have 
the applied field at 45$^{o}$ and the flux is complex with equal real 
and imaginary parts.  As with Fe$_{8}$ there are periodic half-integer 
points at which the splitting is zero but now the maximum splitting 
increases by almost an order of magnitude per period.  The result is 
given in the caption to Fig.~(\ref{Mn}) which compares it with that 
of exact diagonalisation.  Further details will be given 
elsewhere\cite{seb}.  It can also be shown that similar oscillations 
occur when the field is directed along a hard three or four fold axis 
of a cubic material\cite{seb}.

For the basic intermediate spin problem\cite{khare} of a charged 
particle in two dimensions, Fig.~(\ref{interspin}), the relevant group 
is SO(2) which is equivalent to U(1), i.e., is abelian with a single 
operator $S_{z}$, the generator of rotations $e^{iS_{z}\phi}$, where 
the $z$-axis is perpendicular to the circle and where $\phi$ is the 
angle to the $x$-axis.  For a unit circle $S_{z} = p_{\phi}$ the 
momentum conjugate to $\phi$, i.e., $[\phi, p_{\phi}]=i\hbar$.  The 
eigenstates of $S_{z}$ and $p_{\phi}$ are $\psi_{m}(\phi) = {1 \over 
(2\pi)^{1/2}}e^{im\phi}$ but for particles on a circle there is no 
reason to insist on $m$ being an integer reflecting the fact that the 
spectrum of $S_{z}$ for $SO(2)$ is continuous\cite{khare}.  In general 
for this {\it singular gauge} $\psi_{m}(\phi)$ is multi-valued.  A 
convenient single valued function can be defined by taking some 
convention for $\psi_{m}(\phi)$ in the interval $\{0,2\pi\}$ and then 
analytically continuing this function to the interval 
$\{-\infty,+\infty\}$.  For any physical potential $V(\phi) = 
V(\phi+2\pi n)$; $n=0, \pm 1, \pm 2 \ldots$, and implies that only the 
basis states $\psi_{{\alpha_{s}\over 2}+n}(\phi)$; $n=0, \pm 1, \pm 2 
\ldots$, mix.  It follows for a given statistical parameter 
$\alpha_{s}$ the energy spectrum is discrete.  This can be stated 
differently: Since $V(\phi)$ is periodic the solutions obey Floquet's 
(Bloch's) theorem, i.e., $\psi_{k}(\phi) = e^{ik\phi}u_{k}(\phi)$ 
where $u_{k}(\phi)$ is periodic.  The reciprocal lattice vector $K = 
{2\pi/ (a= 2\pi)} = 1$ and the energy $\epsilon(k)=\epsilon(k+nK); \ \ 
n=0, \pm 1, \pm 2 \ldots$ i.e., is periodic.  An examination of the 
Fourier components immediately implies $k= \alpha_{s}/2$.  Given that 
$V(\phi)$ has a point with reflection symmetry the $\psi_{\pm 
k}(\phi)$ are degenerate and the most general solution for a given 
energy is\break
\begin{figure}[b]
\vglue -1.1truecm
\centerline{\epsfig{file=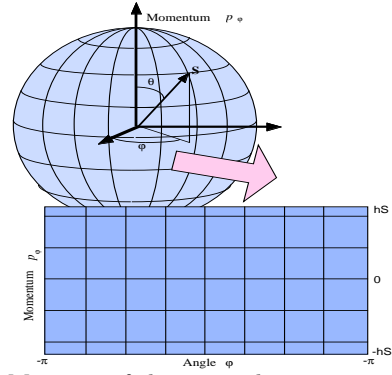,height=2in,width=2.in} } 
\caption[toto]{
Mapping of the unit sphere to a two dimensional 
rectangle
}
\label{sphere}
\end{figure}
\noindent  $A \psi_{k}(\phi) + B 
\psi_{-k}(\phi)$ for which the spectrum of $m = n \pm {\alpha_{s}\over 
2}$, i.e., there are twice the number of observable $S_{z}$ values 
{\it except\/} importantly for the whole and half-integer points when 
$\alpha_{s} = 0$ or $1$.  This more general method of extrapolating 
between whole and half-integer spin which will be found applicable 
here.

In the {\it non-singular gauge\/} the basis set $\psi_{n}(\phi)$ has 
$n = \pm 1, \pm 2 \ldots$ and the charged particle encircles a 
magnetic flux $\Phi = \alpha_{s} \Phi_{0}/2$ where $\Phi_{0}$ is the 
appropriate flux quantum.  For the same $V(\phi)$ there exists an 
eigenstate $u_{k}(\phi)$ with the same energy as in the singular 
gauge, i.e., the energy spectrum are identical and two problems can 
mapped to each other and both reflect intermediate spin.  In the 
non-singular gauge the combination $A u_{k}(\phi) + B u_{-k}(\phi)$ 
comprises a linear combination of states with the flux $\Phi$ in both 
its positive and negative senses.  Evidently the expectation value of 
the flux $\langle \Phi \rangle$ is in general smaller than 
$\alpha_{s} \Phi_{0}/2$.  This possibility of a linear combination of 
different flux states does not seem to have been envisaged in 
connection with intermediate statistics although it would arise if, 
e.g., quantum coherence in SQUIDS could be observed.  It is realized 
in the systems, e.g., Fe$_{8}$, discussed here.

The usual functional integral formulation\cite{review} involves 
coherent state $|\theta, \phi \rangle$ obtained by rotating $|S_{z} = 
S\rangle$ through usual angles $\theta, \phi$.  {\it However\/} the 
$|\theta, \phi \rangle$ are {\it very\/} much over-complete.  In fact 
the states $| \phi \rangle \equiv |\theta={\pi \over 2} , \phi \rangle 
= (2S+1)^{1/2} \sum_{m=-S}^{S} e^{-im\phi}|m \rangle $, where $|m 
\rangle \equiv |S_{z} = m \rangle$, remain over-complete.  If $ | \psi 
\rangle = \sum_{m=-S}^{S} a(m) |m \rangle$ then $\psi(\phi) \equiv 
\langle \phi | \psi \rangle = \sum_{m=-S}^{S} e^{im\phi} a(m)$, i.e., 
is a Fourier transform of the $m$-space wave function $a(m)$.  The 
function $\psi(\phi)$ is defined on the interval $\{-\infty,+\infty\}$ 
and the inverse transform $\psi(m) \equiv {1\over 2\pi} 
\int_{-\infty}^{+\infty}\, d\phi \psi(\phi) = \sum_{m^{\prime}} 
\delta(m - m^{\prime}) a(m^{\prime})$ so that $ | \psi \rangle = 
\int_{-\infty}^{+\infty}\, dm\, \psi(m) |m \rangle$, i.e., $\psi(m)$ 
is a definition of the $m$-space wave function appropriate when $m$ is 
considered as a continuous variable.  Clearly $\hat S_{z} \psi(\phi) = 
\int_{-\infty}^{+\infty}\, dm e^{im\phi} \hbar m \psi(m) = p_{\phi} 
\psi(\phi) $ where $ p_{\phi} = - i \hbar {d \over d \phi}$ is the 
same as that defined earlier.  Also evident is that multiplication by 
$e^{i\phi}$ amounts to a displacement of $m$ by unity and so, e.g., 
$\hat S^{+} = e^{i\phi}[\hbar^{2}S(S+1) - p_{\phi}(p_{\phi}+1)]^{1/2} $.  
These definitions permit an arbitrary spin Hamiltonianto be expressed 
in terms of the cononical $p_{\phi}$ and $\phi$.

In terms of the unit sphere $\phi$ and $p_{\phi}$ might be considered 
as a particular projection in which the axis of quantization plays the 
role of the North-South axis as shown in Fig.~(\ref{sphere}).  In 
these terms, the angle $\phi$ is evidently the longitude while 
$p_{\phi}$ is the projection on the $z$-axis and is therefore the sine 
of the latitude.

It is necessary to carefully formulate the problem in terms of the 
basis set $|m\rangle$, i.e., in $m$-space.  If $|\psi \rangle = 
\sum_{m=-S}^{+S}a(m)|m\rangle$, then in this basis Schr\"odinger's 
equation for Eqn.~(\ref{un}) can be written as ($h_{\ell} = h_{t} =0$):
\begin{eqnarray}
&\left(\epsilon - (2E\hbar^{2} m^{2} - h\hbar m) \right) a(m) \nonumber \\
&= 
{1\over 4}\hbar^{2} (D-E)\Big[
M_{m}^{m+1} M_{m+1}^{m+2} a(m+2) \nonumber \\
&+
M_{m}^{m-1} M_{m-1}^{m-2} a(m-2)\nonumber \\
&+[{M_{m}^{m+1}}^{2}+{M_{m}^{m-1}}^{2}]a(m)
\Big], 
\label{chain}
\end{eqnarray}
where $M_{s}^{t} = \left[ S(S+1) - st \right]^{1/2}$.  The structure 
of this equation is worth noting.  It has the form of {\it two\/} 
tight binding models, one for even $m$ values and the other for odd 
$m$. 

The wave function for the eigenstate $|S_{x} = S\rangle$ is $a^{0}(m) 
= {1\over 2^{S}} \left[{(2S)!  \over (S-m)!(S+m)!}\right]^{1/2}$ and 
$a^{0}(m)$ has to be a solution to Eqn.~(\ref{chain}) with $h=E=0$, 
this corresponding to the absence of tunneling.  Using Stirling's 
approximation this reduces to $a^{0}(m) \propto e^{-m^2/2S}$, however 
the deviations from this approximation are very important.  The 
absence of tunneling when $h=E=0$ is put in evidence by a 
transformation: $a(m) = a^{0}(m) f(m)$.  With the definition $f(\phi) 
= \sum_{m} e^{im\phi} f(m)$ this yields the {\it exact\/} 
Schr\"odinger's equation:
\begin{equation}
	\Big({\cal E} - 2E{p_{\phi}}^{2}\Big) 
f(\phi) =  V(\phi, p_{\phi}) f(\phi)
	\label{cinq}
\end{equation}
with
\begin{eqnarray}
&V(\phi, p_{\phi})	 = {1\over 2} (D-E) \times\nonumber \\
&
\Big[
2\sin^{2}\phi(\hbar^{2} S(S-1)+{p_{\phi}}^{2})
- i \hbar (2S-1)\sin 2\phi p_{\phi}  
\Big],
	\label{six}
\end{eqnarray}
where ${\cal E} = (\epsilon - E^{0}); \ E^{0}=-\hbar^{2} S^{2}(D+E))$.  
This is of the intermediate spin form but with a momentum dependent 
potential, $V(\phi, p_{\phi})$.

A new wave function is defined by $f(\phi) = (1-(\alpha_{s} 
p_{\phi}/2\hbar S)) g(\phi)$.  This admixes, e.g., a little of the 
first excited into the ground state.  If $g(\phi)$ is the solution of
\begin{eqnarray}
	&\Big({\cal E} - 2E\left( p_{\phi}+\hbar {\alpha_{s} \over 2}\right)^{2}
&+\hbar^{2}  E {{\alpha_{s}}^{2} \over 2} \Big) 
g(\phi)	\nonumber \\
&& =  V(\phi, p_{\phi}+\hbar {\alpha_{s} \over 2}) 
g(\phi),
	\label{neuf}
\end{eqnarray}
then for modestly large $S$, $f(\phi) = (1-(\alpha_{s} p_{\phi}/2\hbar 
S)) g(\phi)$ is the solution to
\begin{eqnarray}
	&\Big({\cal E} - 2E{ p_{\phi}}^{2} + 
	&2 \hbar (D+E) {\alpha_{s} \over 2} p_{\phi}
 \Big) 
f(\phi)	 \nonumber \\
&&=  V(\phi, p_{\phi}) 
f(\phi).
	\label{dix}
\end{eqnarray}
The shift in $p_{\phi}$ determines the wave vector: $k = 
{\alpha_{s}/2}$.  In turn the statistical parameter must be chosen to 
give ground state energy determined without tunneling.  Required is 
that $\hbar^{2} E {{\alpha_{s}}^{2} \over 2} = {h^{2} \over 8(D+E)}$, 
so $k= {\alpha_{s} \over 2} =h/2\hbar \sqrt{2E(D+E)}$.  This fixes the 
Zeeman term in Eqn.  (\ref{dix}); $2 \hbar (D+E) {\alpha_{s} \over 2} 
p_{\phi} = \sqrt{(D+E)/2E} h p_{\phi}$.  Since, in general, this is 
larger than $h p_{\phi}$ the solution is a Schr\"odinger cat state, $A 
(1-(\alpha_{s} p_{\phi}/2\hbar S)) u_{k}(\phi) + B (1+(\alpha_{s} 
p_{\phi}/2\hbar S)) u_{-k}(\phi)$ involving a linear combination of 
both bare fields $\pm \sqrt{(D+E)/2E} h$ such that the expectation 
value is equal the physical applied field.  This requires $A = 1 + 
\sqrt{2E/(D+E)}$ and $B = 1 - \sqrt{2E/(D+E)}$.  The total energy for 
level $n$ and wave vector $k$ is,
\begin{equation}
E_{n}(k) = E^{0} - {h^{2} \over 8(D+E)}
+ \left(n+{1 \over 2}\right)\hbar \omega_{0} + \epsilon_{n}(k).
\label{onze}
\end{equation}
The first two terms are purely classical in origin while the third 
corresponds to the harmonic motion about the bottom of the well.  It 
is the last term which reflects the tunneling between wells.

The problem therefore reduces to the determination of 
$\epsilon_{n}(k)$ and when the tunneling approximation is well 
justified this reflects the tight binding approximation.  With only a 
transverse field $h$, the problem has a higher symmetry and potential 
$V(\phi, p_{\phi})$ has a $\phi$ period of $a=\pi$ rather than $2\pi$ 
so the reciprocal space unit vector $K=2$.  For a tight binding model 
the tunneling energy, $ \epsilon_{n}(k) = {\Delta_{0} \over 2} \cos 
\pi k$ where $\Delta_{0}/2$ is the undetermined matrix element for 
tunneling for level $n$.  (The method for determining this is given 
in\cite{seb}.)  There are {\it two\/} solutions which are acceptable 
for $h=0$, namely $k = {\alpha_{s} \over 2} = 0$ corresponding to even 
$m$ values {\it and\/} $k = {\alpha_{s} \over 2} =1$ and odd $m$.  
These two $k$ values correspond to $\epsilon(k) = \pm {\Delta_{0} 
\over 2}$ and so the zero field tunnel splitting is reflected in the 
parameter $\Delta_{0}$.  For finite transverse fields $h$ the 
corresponding $k = {\alpha_{s} \over 2} = (h/2\hbar \sqrt{2E(D+E)})$ 
and $k = 1 + {\alpha_{s} \over 2} = 1+(h/2\hbar \sqrt{2E(D+E)})$ and 
so the tunnel splitting is,
\begin{equation}
\Delta = \Delta_{0} \cos  { \pi h \over 2 \hbar \sqrt{2E(D+E)}},
\label{trieze}
\end{equation}
i.e., of the form mentioned in the introductory remarks.  An equivalent 
result was first obtained some time ago by Garg\cite{garg} and attributed to 
``topological quenching''.

Nowhere in the development has use been made of the assumption that 
$S$ is a whole-integer.  The only difference when $S$ is a 
half-integer is that the values of $m$ are also half-integer and thus 
the solutions correspond to $k = {1\over 2} + {\alpha_{s} \over 2} = 
{1\over 2} + (h/2 \hbar \sqrt{2E(D+E)})$ and $k = - {1\over 2} + 
{\alpha_{s} \over 2} = - {1\over 2} + (h/2\hbar \sqrt{2E(D+E)})$.  
This has the effect of replacing the cosine is replaced by a sine.

\begin{figure}[t!]
\centerline{\epsfig{file=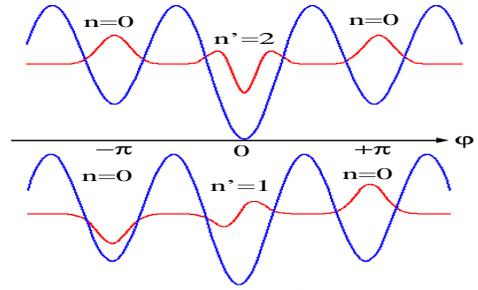,height=1.5in,width=2.5in} } 
\caption[toto]{When, e.g., $n=0$ and $n^{\prime}=2$, both 
wave functions are even there is no sign change.  If $n=0$ and 
$n^{\prime}=1$ the wave must change sign as its $\phi$ argument 
advances by $2\pi$.  }
\label{evenodd}
\end{figure}

Consider first the effect of adding $h_{\ell}$ alone.  This adds a 
potential, i.e., $ V(\phi, p_{\phi}) \Rightarrow V(\phi, p_{\phi}) + 
\hbar S h_{\ell} \cos \phi $ and removes the symmetry between $\phi = 
0$ and $\pi$.  The period of the potential is now the full $a=2\pi$, 
with $K=1$, {\it but\/} with two wells per unit cell.  The harmonic 
levels near $\phi= \pi$ have quantum numbers designated by $n$ and are 
higher than those near $\phi = 0$ with labels $n^{\prime}$.  Ignoring 
tunneling these have energies, $E_{n} =E_{0} - {h^{2} \over 8(D+E)} + 
\hbar n h_{\ell} + \left(n+{1 \over 2}\right)\hbar \omega_{0}, $ and, 
$ E_{n^{\prime}} = E_{0} - {h^{2} \over 8(D+E)} - \hbar n^{\prime} 
h_{\ell} + \left(n^{\prime}+{1 \over 2}\right)\hbar \omega_{0}$.  For 
a given $h$ there is only one solution with $k = {\alpha_{s} \over 2} 
= (h/2 \hbar \sqrt{2E(D+E)})$.  There are now two possibilities (i) 
$E_{n} \ne E_{n^{\prime}}$ and there is a {\it very\/} narrow band 
formed about both these energies.  The energy differences have a {\it 
very\/} small dependence on $k$ and hence $h$, see below.  (ii) When 
$E_{n} \approx E_{n^{\prime}}$ there is resonant tunneling.  There are 
again two cases (iia) when both $n$ and $n^{\prime}$ are odd or even 
and (iib) when one of $n$ and $n^{\prime}$ is odd and the other even.  
As illustrated in Fig.~(\ref{evenodd}) there are two tunneling matrix 
elements one $\Delta_{n,n^{\prime}}$ for the barrier between $\phi = 
0$ and $\pi$ and $\Delta_{n^{\prime},n}$ for that between $\phi = 
-\pi$ and $0$ (or $\pi$ and $2\pi$).  With first $h=(h_{t} =) 0$, 
Schr\"odinger's equation admits real solutions, $\psi(m)$ or 
$\psi(\phi)$, and so these matrix elements can be made real.  For case 
(iia) when the symmetry of the wave functions is the same the 
tunneling matrix elements $\Delta_{n,n^{\prime}}= 
\Delta_{n^{\prime},n} \equiv \Delta_{0}$.  For finite $h$, for the 
Floquet (Bloch) wave function, the phase advances by $2\pi k$ from one 
cell to the next while to agree with the smaller unit cell appropriate 
when $h_{\ell}=0$ the phase should advance by $\pi k$ from one well to 
the next within the cell.  This leads to $\Delta_{n,n^{\prime}}= 
e^{i\pi k} \Delta_{0}$ and $ \Delta_{n^{\prime},n} = e^{-i\pi 
k}\Delta_{0}$.

At this point it is possible to introduce the second transverse field 
$h_{t}$ by slight of hand.  It is observed that if $E$ changes sign 
then the two transverse axes change role and thus $h$ becomes the 
equivalent of $h_{t}$.  If $\cal H$ is considered to be a function of 
a complex parameter $E$ the solution might be analytically continued 
from positive to negative values.  In particular $k = (h/2\hbar 
\sqrt{2E(D+E)})$ becomes pure imaginary, i.e., $k \to ik_{t} = 
i(h_{t}/2\hbar \sqrt{2E(D-E)})$ and it follows that 
$\Delta_{n,n^{\prime}}= e^{i\pi (k+ik_{t})} \Delta_{0}$ with $ 
\Delta_{n^{\prime},n} = e^{-i\pi (k+ik_{t})}\Delta_{0}$.  This 
information can be represented by a two level model:
$$
{\cal H} = (E_{n}-E_{n^{\prime}}) S_{z} + {\Delta_{0}\over 2}
\left[ e^{i\pi (k+ik_{t})} S^{+} + 
e^{-i\pi (k+ik_{t})}S^{-}\right],
$$
where the center of gravity energy ${1\over2} 
(E_{n}+E_{n^{\prime}})$ has been dropped.  Degenerate 
perturbation theory then yields, $ \epsilon_{k} = \pm {1\over2} 
\sqrt{(E_{n}-E_{n^{\prime}})^{2}+ 4{\Delta_{0}}^{2}|\cos 
\pi (k+ik_{t})|^{2}}$ or at resonance $\epsilon_{k} = \pm 
{\Delta_{0}\over 2}|\cos \pi (k+ik_{t})|$, with $k = {h \over 2 \hbar 
\sqrt{2E(D+E)}}$ and $k_{t} = {h_{t} \over 2 \hbar \sqrt{2E(D-E)}}$.  
The $h_{\ell}= h_{t} =0$ result is recovered when $n = n^{\prime}$.  
When for (iib) one wave function\break\vglue -0.5truecm
\noindent changes sign while the other does not 
it is necessarily the case, for $h=h_{t}=0$, that 
$\Delta_{,n^{\prime}}= - \Delta_{n^{\prime},n} \equiv \Delta_{0}$ 
whence everywhere in $ \epsilon_{k}$ the cosine becomes a sine and 
the result is as if the original $S$ was half-integer! This explains 
the alternation observed in experiment\cite{ws}.

Certainly the most striking claim is that the wave function $A 
(1-(\alpha_{s} p_{\phi}/2\hbar S)) u_{k}(\phi) + B (1+(\alpha_{s} 
p_{\phi}/2\hbar S)) u_{-k}(\phi)$ for an intermediate spin system is a 
mixture of a state with the local applied field of magnitude $h_{p} 
=\pm \sqrt{(D+E)/2E} h$, for small fields.  The fields are both 
parallel and anti-parallel to the distant applied field $h$, but 
see\cite{seb}.  (With an equivalent claim for $h_{t}$ which involves 
$\sqrt{(D-E)/2E} h_{t}$.)  {\it Experimentally\/} the existence of 
this superposition is easy to establish.  The transverse field $h$ 
couples to the physical $S_{z}$, i.e., this operator in the 
non-singular gauge, and which has a spectrum with a period $\hbar$ or 
in units of magnetic moment $g \mu_{B}\hbar$ where $g$ is the 
g-factor.  Thus if $M_{z}$ is the measured moment in the direction of 
the transverse field, $m = M_{z}/g \mu_{B}\hbar$, the dimensionless 
magnetization has a period of unity.  This would be the case if the 
state did not consist {\it precisely\/} of a superposition of field 
directions.  Because of the admixture, the measurable period is 
reduced to $ \Delta m \equiv m_{1} = \sqrt{2E \over D+E} \le 1$ which 
is confirmed by exact results but remains to be tested experimentally.


\vskip -15pt

\begin{references}



\bibitem{ws} W. Wernsdorfer and R. Sessoli, Science {\bf 284}, 133 (1999).

\bibitem{barnes} S.E. Barnes, J. Phys. Cond. Matter 10, L665 (1998); 
and cond-mat/9907257.

\bibitem{thomas} L. Thomas, F. Lionti, R. Ballou, R. Sessoli, A. 
Caneschi and B. Barbara, Nature {\bf 383}, 145 (1996); J. R. Friedman 
et al., Phys.  Rev.  Lett.  {\bf 76}, 3820 (1996).

\bibitem{seb} S.E. Barnes, unpublished, cond-mat/00XXXX.

\bibitem{khare} See e.g., A. Khare {\it Fractional statistics and 
quantum field theory}, World Scientific, Singapore (1997).

\bibitem{review} E. M. Chudnovsky and J. Tejada, {\it Macroscopic 
Quantum Tunneling of the Magnetic Moment} (Cambridge University Press, 
1997).

\bibitem{garg}  A. Garg, Europhys. Lett. {\bf 22}, 205 (1993).

\end{references}
\end{document}